\documentstyle[aps,psfig,eqsecnum,preprint]{revtex}
\begin{document}
\tighten
\draft
\preprint{{\small IFIC--00--42}}
\title{The 't Hooft interaction at finite $T$ and $\mu$:\\
Beyond the mean field theory}
\author{Yasuo Umino and Vicente Vento} 
\address{Instituto de F\'{\i}sica Corpuscular -- C.S.I.C. \\
Departamento de F\'{\i}sica Te\`orica, Universitat de Val\`encia \\
E--46100 Burjassot, Val\`encia, Spain}
\date{\today}
\maketitle
\begin{abstract}
We use the N--quantum approach (NQA) to quantum field theory to construct a solution of
the two flavor effective instanton induced `t~Hooft interaction model valid for any 
temperature ($T$) and chemical potential ($\mu$) beyond the mean field theory. 
The model contains only the $\bar{q}q$ channels. In constructing this solution
we calculate the masses, widths and coupling constants of bound $\sigma$ and 
resonant color $\bar{3}$ scalar diquark states. We find that the chemical potential 
induced by the interaction cancel exactly for all values of $T$ in contrast to the 
Nambu--Jona--Lasinio model. Our method can also be extended to include the $qq$ channels
in the model Lagrangian and we discuss how the NQA can be used to study the
properties of bound diquark states at finite $T$ and $\mu$. 
\vskip 0.25cm
\noindent Submitted to Physics Letters B
\end{abstract}

\vfill\eject

\section{Introduction}
\label{intro}
In this letter we apply the N--quantum approach (NQA) to quantum field theory 
\cite{gre94a} to the effective instanton induced `t~Hooft interaction Lagrangian 
\cite{tho96} and construct a non--perturbative solution to the field theory 
beyond the mean field approximation for any temperature ($T$) and chemical
potential ($\mu$).  Developed by Greenberg \cite{gre65}, NQA uses an expansion 
in on--shell asymptotic fields, known as the Haag expansion \cite{haa55}, to solve the 
operator equations of motion. The method is particularly well suited for bound 
state problems since it describes bound states with the same number of kinematical
variables as in non--relativistic theories while maintaining explicit
Lorentz invariance (for theories in free space). A summary of applications of 
NQA can be found in \cite{gre94a}.

Recently we have extended the NQA to finite $T$ and $\mu$ \cite{umi00}. 
Our work was based on the thermal 
field dynamics formalism of Umezawa and collaborators \cite{umezawa}, and was  
motivated by an application of NQA to the BCS theory of superconductivity 
\cite{gre94b}. Temperature and chemical potential were introduced simultaneously 
by applying a thermal Bogoliubov transformation to each of
the fermionic or bosonic field operators in the Haag expansion. At
the same time we defined a thermal vacuum state which is annihilated by
each of the thermal annihilation operators. This takes care of the
doubling of the Hilbert space which characterizes the formalism of thermal field 
dynamics. Furthermore, in the NQA one works only 
in the space of asymptotic field operators and therefore there is no need to
specify the structure of this vacuum state. 

We have successfully tested our formalism on the Nambu--Jona--Lasinio (NJL) model 
\cite{nam61} where the mean field (Hartree--Fock) results at finite $T$ and $\mu$ 
are well established \cite{reviews}. These results were recovered using only the 
first term in the Haag expansion. In this work we take the second order terms in the 
Haag expansion into account to solve the `t~Hooft model beyond the mean field theory. 
Although non--renormalizable and non--confining, the `t~Hooft interaction
incorporates the symmetries of QCD and naturally provides a mechanism for
chiral symmetry breaking and resolves the $U_A(1)$ problem \cite{sch98}. 
As is well known it also supports the formation of various kinds of condensates. 
Among them are the color singlet chiral condensate ($\langle \bar{q}q \rangle$) which 
breaks chiral symmetry dynamically and generates a mass to the fermions and
the color anti--triplet scalar diquark condensate ($\langle qq \rangle$) which 
triggers color superconductivity at finite density \cite{diquark}. 

In this work we only include the $\bar{q}q$ channels in the model Lagrangian. Our solution 
is constructed by calculating the masses, widths and coupling constants of the $\sigma$ 
and {\em resonant} diquark states for any $T$ and $\mu$ both in the Nambu--Goldstone and in 
the Wigner--Weyl phases of the model field theory. In the 
concluding section we discuss how our method can be used to study the 
properties of the {\em bound} diquark states in hot and dense matter. 

\section{Model Lagrangian and the NQA Ansatz}
\label{model}

We start by defining the model Lagrangian which is the 
SU(2)$_{\rm f}$ $\otimes$ SU(3)$_{\rm c}$ version of the `t~Hooft interaction 
\cite{tho96} with massless quarks and a chemical potential term.
\begin{eqnarray}
{\cal L} 
& = & 
i\bar{\Psi} \left( \partial\!\!\!/  -i \mu_0 \gamma_0 \right) \Psi + 
g_0 \Bigl[ (\bar{\Psi} \Psi )^2 + (\bar{\Psi} i \vec{\tau}\gamma_5 \Psi)^2 
- (\bar{\Psi} i\gamma_5 \Psi)^2 - (\bar{\Psi}\vec{\tau} \Psi)^2 \Bigr]
\label{eq:tHooft} \\
& = &
{\cal L}_{\rm NJL} - g_0 \Bigl[ (\bar{\Psi} i\gamma_5 \Psi)^2 
+ (\bar{\Psi}\vec{\tau} \Psi)^2 \Bigr]
\label{eq:tHooft2} 
\end{eqnarray}
In the above Lagrangian $\bar{\Psi} \Psi$,
$\bar{\Psi} i \vec{\tau}\gamma_5 \Psi$, $\bar{\Psi}i\gamma_5 \Psi$ and
$\bar{\Psi}\vec{\tau} \Psi$ terms correspond to the $\bar{q}q$ channels with quantum numbers
of the $\sigma$, $\pi$, $\eta'$ and $a_0$ mesons, respectively. The coupling constant 
$g_0$ has the dimension of $[{\rm mass}]^{-2}$. In
Eq.~(\ref{eq:tHooft2}) the `t~Hooft Lagrangian is expressed as a sum of the
NJL lagrangian and interactions in the $\eta'$ and $a_0$ channels. Note that 
$\mu_0$ is in general not the total chemical
potential of the interacting many body system since the interaction may induce a
contribution to the total chemical potential.  

For our purpose the equation of motion corresponding to Eq.~(\ref{eq:tHooft}) is most
conveniently written in terms of colored quark fields in momentum space. For example,
the charge conjugation symmetric form of the equation of motion for the up quark field 
may be written as
\begin{eqnarray}  
(q\!\!\!/ + \mu_0 \gamma_0)_{\alpha\beta}(u_a)_\beta(q) 
& = &
-2 g_0 \int d^4p_1 d^4p_2 d^4p_3\; \delta^4(q - p_1 - p_2 - p_3) \nonumber \\
&   &
\!\!\!\!\!\!\!\!\!\!\!\!\!\!\!\!\!\!\!\!\!\!\!\!\!\!\!\!\!\!\!\!\!\!\!\!
\otimes \;
\Biggl\{ \left[(\bar{d}_b)_\beta(p_1), (d_b)_\beta(p_2) \right]_- (u_a)_\alpha(p_3)
+ \left[(\bar{d}_b)_\beta(p_1), (d_a)_\alpha(p_2) \right]_- (u_b)_\beta(p_3) \nonumber \\
&   &
\!\!\!\!\!\!\!\!\!\!\!\!\!\!\!\!\!\!\!\!\!\!\!\!\!\!\!\!\!\!\!\!\!\!\!
\!\!\!\!\!\!\!\!\!\!\!\!\!\!\!\!\!\!\!\!\!
+ \;\; 
(\gamma_5)_{\alpha\beta}(\gamma_5)_{\epsilon\rho} \Biggl[
\left[(\bar{d}_b)_\epsilon(p_1), (d_b)_\rho(p_2) \right]_- (u_a)_\beta(p_3)
+ \left[(\bar{d}_b)_\epsilon(p_1), (d_a)_\beta(p_2) \right]_- (u_b)_\rho(p_3)
\Biggr] \Biggr\}
\label{eq:motion}
\end{eqnarray}
where $[A, B]_- = AB - BA$.  $u$ and $d$ are up and down quark fields with color and 
Dirac indices indicated in Roman and Greek letters, respectively. Summation convention 
for repeated indices is implied. 

We proceed to construct an ansatz for the `t~Hooft model by Haag expanding each of 
the quark fields in Eq.~(\ref{eq:motion}) to second order following \cite{gre86}. 
For example, the second order Haag expansion for the up quark field with color index 1 
may be written as
\begin{eqnarray}
(u_1)_\alpha(q)
& = &
(u_1^0)_\alpha(q) + \int d^4k d^4b\; \delta^4(q + k -b)
\Biggl\{ f_{\alpha\beta}(k, b):(u_1^0)_\beta(-k)\sigma^0(b): \nonumber \\
&   &
+ d_{\alpha\beta}(k, b) (C^{-1})_{\beta\epsilon}
:\Bigl[ (\bar{d}_2^0)_\epsilon(-k) D_2^0(b) + (\bar{d}_3^0)_\epsilon(-k)
D_5^0(b)\Bigr]:\Biggr\} 
\label{eqn:haagu1} 
\end{eqnarray}
where the symbol $:\hspace{0.5cm}:$ denotes normal ordering. The asymptotic fields
in the Haag expansion are choosen to be in--coming fields which implies that the 
corresponding propagators obey the retarded boundary conditions. These fields are
 defined as 
$u_a^0(k) = u_a^{\rm IN}(k)\delta(k^2 - m^2)$,
$d_a^0(k) = d_a^{\rm IN}(k)\delta(k^2 - m^2)$,
$\sigma^0(b) = \sigma^{\rm IN}(b)\delta(b^2 - m_\sigma^2)$ and
$D^0_A(b) = D_A^{\rm IN}(b)\delta(b^2 - m_D^2)$. 
In Eq.~(\ref{eqn:haagu1}) $\sigma$
is a color singlet scalar field which transforms as $\bar{\Psi}\Psi$ while
$D_A$ is a color anti--triplet scalar field transforming as 
$\Psi^T C^{-1} \gamma_5 \tau_2 \lambda_A \Psi$ with $A = 2, 5, 7$ and 
$C^{-1} = -C = -i \gamma_2\gamma_0$ being the charge conjugation operator.

In this work we only include the $\sigma$ and $D_A$ fields in the second order Haag 
expansion. Other composite fields may be included 
in the second order terms as shown, for example, in \cite{gre86}. 
We emphasize that $\sigma$ and $D_A$ fields need {\em not} be interpreted as bound
states. As we shall see below, in medium they can be either bound or resonant states. 
Thus, in order to avoid confusion with the bound diquark state \cite{diquark} we shall
henceforth refer the color $\bar{3}$ $qq$ state simply as the $D_A$ state. 

The asymptotic quark mass $m$ will shortly be identified as 
the chiral gap while the $\sigma$ and $D_A$ masses $m_\sigma$ and $m_D$ will be
determined below as functions of $T$ and $\mu$. The Haag amplitudes $f(k, b)$ and 
$d(k, b)$ are interpreted as amplitudes for the creation of the $\sigma$ and $D_A$ 
states, respectively. 
Corresponding expansion for $(\bar{u}_1)_\alpha(q)$ can be found by using the relation
$(\bar{u}_1)(q) = u_1^\dagger(-q)\gamma^0$. Second
order Haag expansions for the remaining quark fields in Eq.~(\ref{eq:motion}) 
can be constructed in a similar manner. 

We use the method developed in \cite{umi00} to introduce temperature and chemical 
potential into Eq.~(\ref{eqn:haagu1}). The on--shell fermionic field $\Psi^0$ is 
extended to finite $T$ and $\mu$ as follows.\footnote{We use the 
conventions and normalizations of Itzykson and Zuber \cite{IZ}.}
\begin{eqnarray}
\Psi^0(q)
& = &
\Psi(q)\delta(q^2 - m^2) \nonumber \\
& = &
\frac{1}{(2\pi)^3}\frac{m}{\omega_q} \sum_{s=\pm} 
\biggl[ b(\vec{q},s) u(\vec{q},s) \delta(q^0-\omega_q) + d^{\dagger}(-\vec{q},s) v(-\vec{q},s) 
\delta(q^0+\omega_q) \biggr]  \nonumber \\
& \rightarrow &
 \frac{1}{(2\pi)^3}\frac{m}{\omega_q} \sum_{s=\pm} \Biggl\{
\biggl[ \alpha_q B(\vec{q},s) - s \beta_q \tilde{B}^{\dagger}(-\vec{q},s)\biggr] 
u(\vec{q},s) \delta(q^0-\omega_q)  \nonumber \\
&    &
\;\;\;\;\;\;\;\;\;\;\;\;\;\;\;\;\;\;\;\; +
\biggl[ \gamma_q D^{\dagger}(-\vec{q},s) - s \delta_q 
\tilde{D}(\vec{q},s) \biggr] v(-\vec{q},s) \delta(q^0+\omega_q) \Biggr\}
\label{eq:FERMITMU}
\end{eqnarray}
Here $B$ and $\tilde{B}^{\dagger}$ are quasi--particle annihilation and
quasi--hole creation operators, respectively, while $\tilde{D}$ and $D^{\dagger}$ are
the annihilation operator for a quasi--anti--hole and creation operator for a 
quasi--anti--particle, respectively. These thermal operators satisfy the fermion 
anti--commutation relations and annihilate the interacting thermal vacuum state 
$|{\cal G}(T,\mu)\rangle$. 
\begin{equation}
B(\vec{q},s) |{\cal G}(T,\mu)\rangle 
=\tilde{B}(\vec{q},s) |{\cal G}(T,\mu)\rangle 
= D(\vec{q},s) |{\cal G}(T,\mu)\rangle 
=\tilde{D}(\vec{q},s) |{\cal G}(T,\mu)\rangle 
= 0
\label{eq:AVAC1}
\end{equation}
The $T$ and $\mu$ dependent coefficients of the 
thermal Bogoliubov transformation are given by $\alpha_q = \sqrt{1-n_q^-}$,
$\beta_q = \sqrt{n_q^-}$, $\gamma_q = \sqrt{1-n_q^+}$ and
$\delta_q = \sqrt{n_q^+}$ where 
$n_q^{\pm}= [e^{(\omega_q \pm \mu)/(k_B T)}+1]^{-1}$ are the Fermi
distribution functions.

The extension to finite $T$ and $\mu$ for the on--shell bosonic field $\phi^0$ 
proceeds through an analogous path. 
\begin{eqnarray}
\phi^0(q)
& = &
\phi(q)\delta(q^2 - m^2) \nonumber \\
& = &
\frac{1}{2 \omega_q} \biggl[ a(\vec{q}\:) \delta(q^0-\omega_q) 
+ a^{\dagger}(-\vec{q}\:) \delta(q^0+\omega_q) \biggr]  \nonumber \\
& \rightarrow &
\frac{1}{2 \omega_q} \Biggl\{
\biggl[ c_q A(\vec{q}\:) + d_q \tilde{A}^{\dagger}(\vec{q}\:) \biggr] 
\delta(q^0-\omega_q) \nonumber \\
&    &
\;\;\;\;\;\;\;\;\;\;\;\;\;\;\;\;\;\;\;\; + 
\biggl[ d_{q} \tilde{A}(-\vec{q}\:) + c_{q} A^{\dagger}(-\vec{q}\:) \biggr] 
\delta(q^0+\omega_q) \Biggr\}
\label{eq:BOSETMU}
\end{eqnarray}
The coefficients of the bosonic thermal Bogoliubov
transformation are given by $c_q = \sqrt{1 + d_q^2}$ and $d_q = \sqrt{n_q^B}$ where 
$n_q^B = [e^{(\omega_q- \mu)/(k_B T)}-1]^{-1}$ is the Bose distribution function. 
The thermal operators now obey the boson commutation relations and annihilate the 
thermal vaccum as in Eq.~(\ref{eq:AVAC1}),
\begin{equation}
A(\vec{q}\:) |{\cal G}(T,\mu)\rangle
=\tilde{A}(\vec{q}\:) |{\cal G}(T,\mu)\rangle 
= 0
\label{eq:AVAC2}
\end{equation}

Thus our ansatz for the model Lagrangian Eq.~(\ref{eq:tHooft}) at finite 
$T$ and $\mu$ is constructed in three steps. First, each of the colored quark 
fields in the equations of motion are Haag expanded to second order as in 
Eq.~(\ref{eqn:haagu1}). Second, each of the asymptotic quark fields in the Haag
expansion are subjected to a thermal Bogoliubov transformation of type shown in
Eq.~(\ref{eq:FERMITMU}). Finally, each of the asymptotic bosonic fields 
$\sigma^0$ and $D_A^0$ are subjected to a thermal Bogoliubov transformation of type 
shown in Eq.~(\ref{eq:BOSETMU}). The basic idea of NQA is to use the
operator equation of motion Eq.~(\ref{eq:motion}) to solve for the Haag amplitudes 
$f(k, b)$ and $d(k, b)$ and thus construct a solution to the field theory defined 
by Eq.~(\ref{eq:tHooft}).

\section{The Gap and Self--consistency Equations}
\label{equations}

We now derive the gap equation and the self--consistency equations for the Haag
amplitudes using the operator equation of motion in the one--loop 
approximation. This approximation corresponds
to keeping only those terms involving a single contraction when renormal ordering
the field operators with respect to the interacting vacuum $|{\cal G}(T,\mu)\rangle$. 
We begin by inserting the first order terms in the Haag expansion of
the quark fields in Eq.~(\ref{eq:motion}) which is equivalent to invoking 
the mean field approximation. After renormal ordering
and keeping only the linear terms in the field operators we obtain the
following equation
\begin{equation}
m = 
(24 + 4) g_0 \int d^3\bar{p}\; \frac{m}{\omega_p} 
\left( 1- \beta^2_p - \delta^2_p\right)
\label{eqn:m}  
\end{equation}
where $d^3\!\bar{p} \equiv d^3\!p/(2\pi)^3$. The factors of 24 and 4 represent 
contributions from the Hartree and Fock terms, respectively. In our approach
both of these contributions arise naturally from the equation of motion. This gap 
equation determines the asymptotic quark mass $m$ which is interpreted as the chiral gap. 
As in the NJL model when $m(T, \mu) = 0$ the model is in the Wigner--Weyl phase 
while if $m(T,\mu) \neq 0$ then it is in the Nambu--Goldstone phase.

We find that the chemical potentials induced by the scalar and pseudoscalar
Fock terms in the `t~Hooft interaction cancel exactly.\footnote{They are
identified as coefficients of the $\gamma_0$ operator.} Therefore the total chemical
potential, which appears in the Fermi and Bose distribution functions, is simply the 
bare chemical potential $ \mu_0$. 
This situation is in contrast with the NJL model where in the mean field
approximation the Fock term induces a chemical potential proportional to the quark 
number density making it necessary to solve the gap equation self--consistently with 
the total chemical potential \cite{asa89}. As a result, it becomes impossible to clearly
define critical points in the $T-\mu$ plane \cite{asa89}. Eq~(\ref{eq:tHooft2}) shows
that the `t~Hooft Lagrangian can be written as a sum of the NJL Lagrangian and interactions
in the $a_0$ and the $\eta'$ channels. Thus the addition of these channels leads to the 
cancellation of the chemical potential induced by the $\sigma$ and the $\pi$ channels.
This is one important difference between the NJL and the `t~Hooft model at
finite $\mu$ in the mean field approximation.

To go beyond the mean field theory we add the second order terms 
in the Haag expansion to our ansatz for the quark fields. Then, in addition to 
Eq.~(\ref{eqn:m}), we obtain the following self--consistency
equations for the Haag amplitudes from the equation of motion in the one--loop
approximation
\begin{eqnarray}
\left( b\!\!\!/ - k\!\!\!/ - m \right) f(k, b) 
& = &
\nonumber \\
&   &
\!\!\!\!\!\!\!\!\!\!\!\!\!\!\!\!\!\!\!\!\!\!\!\!\!\!\!\!\!\!\!\!\!\!
\!\!\!\!\!\!\!\!\!\!\!\!\!\!\!\!\!\!\!\!\!\!
2 g_0 \int d^3\bar{p}\; \frac{m}{\omega_p}
\Biggl\{ \left( 1- 2\beta^2_p \right)
\Biggl[ f(p^-, b) + \bar{f}(p^+, b) + \gamma_5 f(p^-, b)\gamma_5 
+ \gamma_5\bar{f}(p^+, b)\gamma_5 
\nonumber \\
&   &
\!\!\!\!\!\!\!\!\!\!\!\!\!\!\!\!\!\!\!\!\!\!\!\!
\!\!\!\!\!\!\!\!\!\!\!\!\!\!\!\!\!\!\!\!\!\!
+ \;\; 3 {\rm Tr}\Bigl( f(p^-, b) + \bar{f}(p^+, b) \Bigr) 
+ 3 {\rm Tr}\Bigl( \gamma_5 f(p^-, b) + \gamma_5\bar{f}(p^+, b) \Bigr)\gamma_5 \Biggr]
\nonumber \\
&   &
\!\!\!\!\!\!\!\!\!\!\!\!\!\!\!\!\!\!\!\!\!\!\!\!\!\!\!\!\!
+ \;\left( 1- 2\delta^2_p \right) 
\Biggl[ f(p^+, b) + \bar{f}(p^-, b) + \gamma_5 f(p^+, b)\gamma_5 
+ \gamma_5\bar{f}(p^-, b)\gamma_5 
\nonumber \\
&   &
\!\!\!\!\!\!\!\!\!\!\!\!\!\!\!\!\!\!\!\!\!\!\!\!
\!\!\!\!\!\!\!\!\!\!\!\!\!\!\!\!\!\!\!\!\!\!
+  \;\; 3 {\rm Tr}\Bigl( f(p^+, b) + \bar{f}(p^-, b) \Bigr) 
+ 3 {\rm Tr}\Bigl( \gamma_5 f(p^+, b) + \gamma_5\bar{f}(p^-, b) \Bigr)\gamma_5 
\Biggr]\Biggr\}
\label{eqn:f} 
\end{eqnarray}
and
\begin{eqnarray}
\left( b\!\!\!/ - k\!\!\!/ - m \right) d(k, b) 
& = &
\nonumber \\
&   &
\!\!\!\!\!\!\!\!\!\!\!\!\!\!\!\!\!\!\!\!\!\!\!\!\!\!\!\!\!\!\!\!\!\!
-4 g_0 \int d^3\bar{p}\; \frac{m}{\omega_p}
\Biggl\{ \left( 1- 2\beta^2_p \right)
\Biggl[d(p^+, b) + \gamma_5 d(p^+, b)\gamma_5 
\nonumber \\
&   & 
+ \;\;C^{-1T} d^{T}(p^+, b) C^{T} 
+ \gamma_5 C^{-1T} d^{T}(p^+, b) C^{T} \gamma_5\Biggr]
\nonumber \\
&   &
+ \;\left( 1- 2\delta^2_p \right)
\Biggl[d(p^-, b) + \gamma_5 d(p^-, b)\gamma_5 
\nonumber \\
&   & 
+ \;\; C^{-1T} d^{T}(p^-, b) C^{T} 
+ \gamma_5 C^{-1T} d^{T}(p^-, b) C^{T} \gamma_5 \Biggr]\Biggr\}
\label{eqn:d}
\end{eqnarray}
In the above expressions $p^\pm = (\pm\omega_p, \vec{p\:})$,
$\bar{f}(p, b) = \gamma_0 f^\dagger(-p, -b) \gamma_0$, $b^2 = m^2_{\sigma}$
for Eq.~(\ref{eqn:f}) and $b^2 = m^2_D$ for Eq.~(\ref{eqn:d}). These three
dimensional linear integral equations for $f(k, b)$ and $d(k, b)$ can be solved 
exactly.

\section{The $\sigma$ Amplitude}

In the broken symmetry phase where $m \neq 0$, the discrete $C$, $P$ and $T$ 
transformations constrain the structure of the $\sigma$ amplitude $f(k, b)$ to be
\begin{equation}
f^{\rm NG}(k, b) = \frac{1}{b\!\!\!/ - k\!\!\!/ - m} \left( A + Bb\!\!\!/ \right) 
\frac{(-k\!\!\!/ + m)}{2m}
\label{eq:fsoln}
\end{equation}
where $A$ and $B$ are real and constant diagonal matrices. The superscript NG
indicates that we are in the Nambu--Goldstone phase of the model. The first factor in
Eq.~(\ref{eq:fsoln}) is the quark propagator obeying the retarded boundary
condition. At finite $\mu$ it is necessary to choose a reference frame since Lorentz
invariance is lost at finite density. We shall work in the rest frame of the
bound states which is the appropriate frame to calculate their masses,
widths and coupling constants as will be shown shortly.

Upon inserting Eq.~(\ref{eq:fsoln}) into the self--consistency equation for
$f(k, b)$ we find the following relation between $A$ and $B$ 
\begin{equation}
\left(A + Bb\!\!\!/\right) \left(-k\!\!\!/ + m \right) = \left(I_f A + 2m B \right)
\left(-k\!\!\!/ + m \right)
\label{eq:fsoln0}
\end{equation}
where
\begin{equation}
I_f= 56 g_0  \int d^4\bar{p}\; \delta(p^2 - m^2) 
\frac{2m^2 - b\cdot p}{(b - p + i\epsilon^0)^2 - m^2}
\left( 1- \beta^2_p - \delta^2_p \right)
\end{equation}
with $b_\mu = (\pm m_\sigma, \vec{0}\;)$. Following \cite{gre86} we trace 
Eq.~(\ref{eq:fsoln0}) after multiplying by 1 and
$b\!\!\!/$ to obtain two homogeneous equations for $A$ and $B$. The
solution for this set of equations exists when $I_f = 1$ which
implies
\begin{equation}
f^{\rm NG}(k, b) = \frac{\left( A + 2mB \right)}{b\!\!\!/ - k\!\!\!/ - m}  
\frac{(-k\!\!\!/ + m)}{2m}
\label{eq:fsoln1}
\end{equation}
It remains to determine the $T$ and $\mu$ dependent normalization constant $A + 2mB$.

From the structure of the second order Haag expansion Eq.~(\ref{eqn:haagu1}) we see that
this normalization constant corresponds to the $\sigma qq$ coupling
constant $g_{\sigma qq}$. This coupling constant, as well as the $\sigma$
mass and its width, can be determined by rewriting the condition 
$I_f = 1$ as
\begin{equation}
0 = 1 - 2g_0 \Biggl[ 28\int d^4\bar{p}\; \delta(p^2 - m^2) 
\frac{2m^2 - b\cdot p}{(b - p + i\epsilon^0)^2 - m^2}
\left( 1- \beta^2_p - \delta^2_p \right) \Biggr]
\label{eq:sigpf}
\end{equation}
In the standard treatment of the NJL model the $\sigma$ mass is determined
by the lowest pole of the $\sigma$ propagator in the random phase
approximation (RPA) at zero three momentum. This pole is a solution of the equation 
\begin{equation}
0 = 1 - 2G \Pi_\sigma(m_\sigma, \vec{0}\;)
\end{equation}
where $G$ is the coupling constant in the NJL Lagrangian and $\Pi_\sigma$ is the
$\sigma$ polarization function \cite{reviews}. Since we are working in the
rest frame of the $\sigma$ we can identify the term in square brackets in 
Eq.~(\ref{eq:sigpf}) as the $\sigma$ polarization function in the broken symmetry 
phase
\begin{equation}
\Pi^{\rm NG}_\sigma(b) = 28\int d^4\bar{p}\; \delta(p^2 - m^2) 
\frac{2m^2 - b\cdot p}{(b - p + i\epsilon^0)^2 - m^2}
\left( 1- \beta^2_p - \delta^2_p \right)
\label{eq:spol}
\end{equation}

For $b_\mu = (b_0, \vec{0}\;)$ and $b_0^2 \geq 4m^2$, the real and imaginary parts of 
Eq.~(\ref{eq:spol}) are given by\footnote{The real part of
$\Pi^{\rm NG}_\sigma$ is most easily determined by calculating the imaginary part and 
relating the result to the real part as shown in \cite{kallen}.}
\begin{eqnarray}
{\rm Re}\; \Pi^{\rm NG}_\sigma(b_0) 
& = &
\frac{7}{\pi^2} {\rm P}\; \int^\infty_m dx\; \frac{(x^2 - m^2)^{3/2}}{x^2 -
\frac{1}{4}b_0^2} \Biggl\{ 1 - \left[ e^{(x-\mu)/(k_B T)}\right]^{-1}
- \left[ e^{(x+\mu)/(k_B T)}\right]^{-1} \Biggr\}
\label{eq:respol} \\
{\rm Im}\; \Pi^{\rm NG}_\sigma(b_0) 
& = &
\frac{7}{\pi}\frac{1}{b_0}\left[\frac{1}{4}b_0^2-m^2\right]^{3/2}
\Biggl\{ 1 - \left[ e^{(\frac{|b_0|}{2}-\mu)/(k_B T)}\right]^{-1}
- \left[ e^{(\frac{|b_0|}{2}+\mu)/(k_B T)}\right]^{-1} \Biggr\} 
\label{eq:imspol} 
\end{eqnarray}
The real part of $\Pi^{\rm NG}_\sigma$ is
equivalent to the one obtained by Asakawa and Yazaki in Eq.~(30) of \cite{asa89} 
except for the factor in front of the principal value integral. They obtain a 
factor of $\frac{13}{2}$ whereas our factor is 7.\footnote{Note that 
their coupling constant $G$ is normalized to $G = \frac{4N_{\rm c} +1}{4N_{\rm c}} g$ 
where $g$ is the coupling constant appearing in the NJL Lagrangian.} Note that 
${\rm Re}\;\Pi^{\rm NG}_\sigma$ is a function of 
$b_0^2$ and is independent of the choice of the $b^\pm \equiv (\pm b_0, \vec{0}\:)$ frames. 

The $\sigma$ mass can be obtained directly from Eq.~(\ref{eq:sigpf}) by
noting that the gap equation Eq.~(\ref{eqn:m}) can be rewritten as
\begin{equation}
1 = 28g_0 \int d^4\bar{p}\; \delta(p^2 - m^2) \frac{m_\sigma^2 - 2 b\cdot p}{(b-p)^2 - m^2}
\left( 1- \beta^2_p - \delta^2_p \right) 
\end{equation}
Using this expression in Eq.~(\ref{eq:sigpf}) implies $m_\sigma = 2m$ for all $T$ and $\mu$. 
Moreover, since the width of the $\sigma$ state is given by
\begin{equation}
\Gamma_\sigma = \frac{1}{m_\sigma}|{\rm Im}\;\Pi_\sigma|
\label{eq:widtheq}
\end{equation}
it follows immediately from Eq.~(\ref{eq:imspol}) that the $\sigma$ has vanishing width
in the broken symmetry phase. 

Finally, in the RPA the $\sigma qq$ coupling constant is given by \cite{reviews}
\begin{equation}
g^2_{\sigma qq} = \left[ \frac{\partial {\rm Re}\; \Pi_\sigma}{\partial
b^2}\right]^{-1}_{b^2 = m_\sigma^2}
\label{eq:coupling}
\end{equation}
which fixes the normalization of the $\sigma$ amplitude in the broken
symmetry phase. Thus, we have solved for the Haag amplitude of the $\sigma$
state which can be written as
\begin{equation}
f^{\rm NG}(k, b) = \frac{g_\sigma qq}{b\!\!\!/ - k\!\!\!/ - m}  
\frac{(-k\!\!\!/ + m)}{2m}
\label{eq:fsoln3}
\end{equation}
and obtained its mass, width and coupling constant. In the broken symmetry
phase the $\sigma$ is a stable bound state with zero width and a mass of exactly twice 
the chiral gap for all values of $T$ and $\mu$. 
This conclusion is in agreement with the one obtained 
by Zhuang, H\"{u}fner and Klevansky in the NJL model \cite{zhu94}.

To obtain the $\sigma$ amplitude in the chirally symmetric phase we repeat the
same exercise elucidated above but with $m = 0$. The structure of the Haag
amplitude now becomes
\begin{equation}
f^{\rm BW}(k, b) = -\frac{g_\sigma qq}{b\!\!\!/ - k\!\!\!/} k\!\!\!/ 
\label{eq:fsoln4}
\end{equation}
while the real and imaginary parts of the extracted $\sigma$ polarization
function are found to be
\begin{eqnarray}
{\rm Re}\;\Pi_\sigma^{\rm BW}(b)
& = & 
\frac{7}{\pi^2} {\rm P}\; \int^\infty_0 dx\; 
\frac{x^3}{x^2 - \frac{1}{4}m_\sigma^2} 
\Biggl\{ 1 - \left[ e^{(x-P_F)/(k_B T)}\right]^{-1}
- \left[ e^{(x+P_F)/(k_B T)}\right]^{-1} \Biggr\}
\\
{\rm Im}\;\Pi_\sigma^{\rm BW}(b)
& = &
\frac{7}{8\pi} 
\Biggl\{ 1 - \left[ e^{(\frac{1}{2}m_\sigma-P_F)/(k_B T)}\right]^{-1}
- \left[ e^{(\frac{1}{2}m_\sigma+P_F)/(k_B T)}\right]^{-1} \Biggr\}
\end{eqnarray}
where $P_F$ is the Fermi momentum. In this case the $\sigma$ mass must be determined 
by solving the equation 
\begin{equation}
0 = 1 - 2g_0 {\rm Re}\Pi_\sigma^{\rm BW}(b)
\end{equation}
for each $T$ and  $P_F$. This mass determines the width and coupling constant through
Eqs.~(\ref{eq:widtheq}) and (\ref{eq:coupling}).

The $\sigma$ state in the Wigner--Weyl phase is in general not a stable
bound state. For $T \neq 0$ it is a resonant state with a finite width which
can decay into a quark--anti--quark pair. However, when $T = 0$ and
$P_F \neq 0$ the decay width vanishes for $m_\sigma < 2P_F$
and the $\sigma$ becomes a stable bound state. This is a consequence of Pauli 
blocking which allows the $\sigma$ to decay only when $m_\sigma \geq 2P_F$ \cite{zhu94}.

\section{The $D_A$ Amplitude}

Determination of the $D_A$ amplitude, mass, widths and coupling constant is
analogous to that of the $\sigma$ case. In the Nambu--Goldstone phase the 
$D_A$ amplitude in the broken symmetry phase is given by
\begin{equation}
d^{\rm NG}(k, b) = \frac{g_{Dqq}^{\rm NG}}{b\!\!\!/ - k\!\!\!/ - m} i\gamma_5 
\frac{(-k\!\!\!/ + m)}{2m}
\label{eq:dsoln3}
\end{equation}
where $g_{Dqq}$ is the $D_A$ coupling constant. The extracted polarization 
function is
\begin{equation}
\Pi^{\rm NG}_D(b) = \frac{1}{\pi^3} \int d^4\bar{p}\; \delta(p^2 - m^2) 
\frac{-b\cdot p}{(b - p + i\epsilon^0)^2 - m^2} 
\left[ 1 - 2\beta_p^2 \theta(p_0) -  2\delta_p^2 \theta(-p_0) \right]
\label{eq:dpol1}
\end{equation}
Similarly, the $D_A$ amplitude and the polarization function in the
Wigner--Weyl phase are
\begin{equation}
d^{\rm BW}(k, b) = -\frac{g_{Dqq}^{\rm BW}}{b\!\!\!/ - k\!\!\!/} 
i\gamma_5 k\!\!\!/
\label{eq:dsoln4}
\end{equation}
and  
\begin{eqnarray}
\Pi^{\rm BW}_D(b) 
& = &
\frac{1}{\pi^3} \int d^4\bar{p}\; \delta(p^2) 
\frac{-b\cdot p}{(b - p + i\epsilon^0)^2} 
\nonumber\\
&   &
\;\;\;\;\;\;\;
\otimes
\left\{ 1 - 2\left[e^{(p-\mu)/(k_B T)}+1\right]^{-1} \theta(p_0) 
-  2\left[e^{(p+\mu)/(k_B T)}+1\right]^{-1} \theta(-p_0) \right\}
\label{eq:dpol2}
\end{eqnarray}

Note that the $D_A$ polarization functions Eq.~(\ref{eq:dpol1}) and (\ref{eq:dpol2}) 
are {\em not} symmetric under the
interchange of particle and anti--particle Fermi distribution functions
due to the presence of the Heaviside functions $\theta(p_0)$
and $\theta(-p_0)$. As a consequence there are two possible branches of the 
$D_A$ state. For example the real part of $\Pi^{\rm NG}_D$ on the positive mass
shell $b^+ = (+m_D, \vec{0}\,)$ is 
\begin{eqnarray}
{\rm Re}\Pi^{\rm NG}_D(b^+)
& = &
\frac{2}{\pi^2} \Biggl\{ {\rm P}\int_m^\infty dx\; 
\frac{x^2\sqrt{x^2-m^2}}{x^2 - \frac{1}{4}m_D^2}
+2 {\rm P}\int_m^\infty dx\; 
\frac{x\sqrt{x^2-m^2}}{m_D - 2x} \left[e^{(x-\mu)/(k_B T)}+1\right]^{-1}
\nonumber \\
&   &
\;\;\;\;\;\;\;\;\;\;\;\;
-2 \int_m^\infty dx\; 
\frac{x\sqrt{x^2-m^2}}{m_D + 2x}\left[e^{(x+\mu)/(k_B T)}+1\right]^{-1}
\Biggr\} 
\end{eqnarray}
while on the negative mass shell $b^- = (-m_D, \vec{0}\,)$ the corresponding result is
\begin{eqnarray}
{\rm Re}\Pi^{\rm NG}_D(b^-)
& = &
\frac{2}{\pi^2} \Biggl\{ {\rm P}\int_m^\infty dx\; 
\frac{x^2\sqrt{x^2-m^2}}{x^2 - \frac{1}{4}m_D^2}
-2\int_m^\infty dx\; 
\frac{x\sqrt{x^2-m^2}}{m_D + 2x} \left[e^{(x-\mu)/(k_B T)}+1\right]^{-1}
\nonumber \\
&   &
\;\;\;\;\;\;\;\;\;\;\;\;
+ 2 {\rm P}\int_m^\infty dx\; 
\frac{x\sqrt{x^2-m^2}}{m_D - 2x}\left[e^{(x+\mu)/(k_B T)}+1\right]^{-1}
\Biggr\} 
\end{eqnarray}
The origin of the two branches is due to the unequal treatment of particles and 
anti--particles at finite density. When $\mu = 0$, the particles and anti--particles 
are described by the same Fermi distribution function and thus the $D_A$ polarization 
functions become invariant under the $b^+$ and $b^-$ interchange leading to a single 
branch. 

\section{$\sigma$ and $D_A$ Masses, Widths and Coupling Constants}
\label{results}

We present results for the $\sigma$ and $D_A$ masses, widths and coupling
constants for zero temperature and finite density. Results for other combinations
of $T$ and $\mu$ will be given elsewhere together with relevant technical details
of this work. We use the non--covariant regularization and simply cut off the
divergent integrals using a cut off parameter $\Lambda$. The other input
parameter is the coupling constant $g_0$. We arbitrary choose $\Lambda$ = 0.61 GeV and
$g_0$ = 5.01 GeV$^{-2}$ which yields a dynamical quark mass of
$m$ = 0.305 GeV $\approx$ $\frac{1}{3}M_N$ at $(T, \mu) = (0, 0)$, 
where $M_N$ is the nucleon  mass.\footnote{These are the same parameters used in 
\cite{umi00}.} 

In Figure~1a we plot the $\sigma$ mass as a function of Fermi
momentum. In the broken symmetry phase of the model $m_\sigma = 2m$ where $m$ is
dynamical quark mass. We see that $m_\sigma$ decreases continuously with
increasing $P_F$ and vanishes at the critical Fermi momentum $(P_F)_c$ of
about 0.3 GeV. The phase transition is second order like
in the NJL model. Beyond $(P_F)_c$ the $\sigma$ mass starts to increase
monotonically and by $P_F$ = 0.44 GeV its value is equal to the value in free
space. For all values of $P_F$ we find that $m_\sigma < 2 P_F$ and therefore
it has vanishing width.  Hence we find the $\sigma$ to be a stable bound
state at finite density and zero temperature in this model.

The behaviour of the density dependent $\sigma$ mass is reflected in the
$\sigma qq$ coupling constant shown in Figure~1b as a function of $P_F$.
When $P_F = 0$ the value of $g^2_{\sigma qq}/4\pi$ is about 0.342 and it
continuously increases with density until the chiral symmetry restoration point
where it assumes its maximum value of 0.636. Thus as the $\sigma$ mass
decreases its coupling constant increases indicating the strengthening of the
$\bar{q}q$ binding. Above $(P_F)_c$, the coupling constant starts to
decrease with the increasing $\sigma$ mass. However, when the $\sigma$
mass becomes comparable to its value in free space the coupling constant is
about 50\% larger than the free space value. Hence the $\sigma$ mass is much
more sensitive to changes in its coupling constant in the Wigner--Weyl
phase than in the Nambu--Goldstone phase.

We find that there are no solutions to the equation determining the $D_A$
mass in the $b^-$ frame with our choice of input parameters. The reason is
the rather small value of the regulator $\Lambda$. We can find
solutions for the $D_A$ mass in this branch if we cut off the divergent
integrals in the $D_A$ polarization function independently of $\Lambda$. 
However, we choose to work with only one regulator as in previous works of
the NJL model at finite $T$ and $\mu$ \cite{reviews}. With a single regulator
we find solutions for the diquark mass in the $b^+$ frame but only for 
$P_F$ $>$ 0.06 GeV. Thus there are no $D_A$ states
in free space and for $T=0$ its formation occurs only above a critical density
the value of which depends on the input parameters. This implies that below
this critical density there is no contribution to the solution of the `t~Hooft 
model from the $D_A$ term in the Haag expansion. We also find solutions for the 
$D_A$ width indicating that this is a resonant state in contrast to the 
$\sigma$ state. The $D_A$ mass and its width are shown in Figure~2a.

In the broken symmetry mode the $D_A$ mass peaks around $P_F$ = 0.2 GeV and
decreases monotonically until the phase transition point. Beyond this point
it increases almost linearly as in the $\sigma$ case. We find that
$m_D > m_\sigma$ for all values of density in which there is a resonant $D_A$ state.
However, the $D_A$ coupling constant is one order of magnitude smaller
than that of the $\sigma$ state as shown in Figure~2b. In fact, $g_{Dqq}$
does not become appreciable until $P_F$ $\approx$ 0.2 GeV and continuously
increases with increasing density. Since $g_{Dqq}$ normalizes the $D_A$
amplitude this means that the contribution from the $D_A$ term in the Haag
expansion becomes important with increasing density but it will
always be a correction to the dominant $\sigma$ term.

\section{Conclusion}
\label{concl}
In this work we have constructed a solution to the `t~Hooft interaction
Lagrangian valid for finite temperature and chemical potential beyond the
mean field approximation. To our knowledge this is the first time that such
a solution has been presented in the literature. In order to construct this solution
we have used the NQA to quantum field theory involving the second order Haag
expansion in the one--loop approximation. We have only included the color 1 
$\sigma$ and color $\bar{3}$ $D_A$ composite fields in our ansatz to the `t~Hooft 
model. An improved solution may be obtained by adding all possible channels, 
both color singlet and non--singlet, allowed by the model Lagrangian. 

Our solution was obtained by solving the self--consistency equations for the
$\sigma$ and $D_A$ Haag amplitudes. This has been accomplished by calculating the 
masses, widths and coupling constants of the $\sigma$ and the $D_A$ states. At finite 
density and zero temperature we find that the $\sigma$ is a stable bound state with a 
vanishing width while the $D_A$ is a resonant state with a finite width. The latter state
is resonant because we have only included the $\bar{q}q$ channels in our model Lagrangian. 
Thus our model provides no mechanism to create a bound $D_A$ state. These properties 
of the $\sigma$ and the $D_A$ states can also be calculated using the standard 
Hartree--Fock--RPA approach. We stress that the advantage of using the NQA over the 
Hartree--Fock--RPA approach is that one can explicitly obtain a solution to the model 
field theory.

It would be interesting to apply our method to study the properties of a bound diquark 
state \cite{diquark} at finite $T$ and $\mu$. This may be done by adding a color 
$\bar{3}$ $qq$ channel
to our model and repeat the calculation presented here using the same ansatz. A vanishing
diquark width would indicate a bound state just as in the case of the $\sigma$ shown in
Figure~1a. In addition we can determine the color superconducting gap by calculating the
off--diagonal Hamiltonian and demanding that it vanishes. We hope to report on the outcome 
of this investigation in the near future.

\acknowledgements
Y.~U. would like to thank G.~Ripka for a very useful discussion on this work. 
This work is supported in part by SEUIYD--PB97--1227. Y.~U. is also supported 
by the fellowship of Ministerio de Education y 
Ciencia de Espa\~{n}a under the auspices of the program "Estancias 
Temporales de Cientificos y Tecn\'{o}logos Extranjeros en Espa\~{n}a".
%
% BIBLIOGRAPHY

%
\vfill\eject
%
% FIGURES
%
\centerline{FIGURE CAPTIONS}
\vskip 1cm
FIGURE~1. \hspace{0.1 cm}
$\sigma$ mass and $\sigma qq$ coupling constant as
functions of Fermi momentum $P_F$ for $T= 0$. Input parameters are $\Lambda$
= 0.61 GeV and $g_0$ = 5.01 GeV$^{-2}$. The critical Fermi momentum is
$(P_F)_{\rm C}$ = 0.301 GeV. (a) $m_\sigma$ vs. $P_F$. The $\sigma$ is a
stable bound state and has vanishing width. (b) $g^2_{\sigma qq}/4\pi$.
\vskip 0.75cm
FIGURE~2. \hspace{0.1 cm}
Color $\bar{3}$ scalar diquark mass and $Dqq$ coupling constant evaluated in the frame
$b_\mu = (+m_D, \vec{0}\;)$ as functions of Fermi momentum $P_F$ for $T= 0$. Input
parameters are the same as in Figure~1. (a) $m_D$ vs $P_F$. The diquark
state does not exists for $P_F$ $<$ 0.06 GeV. It is a resonant state and its
width $\Gamma$ is also shown in the figure bounded by the dashed line. 
(b) $g^2_{Dqq}/4\pi$.  
\vfill\eject
\centerline{Figure 1}
\psfig{figure=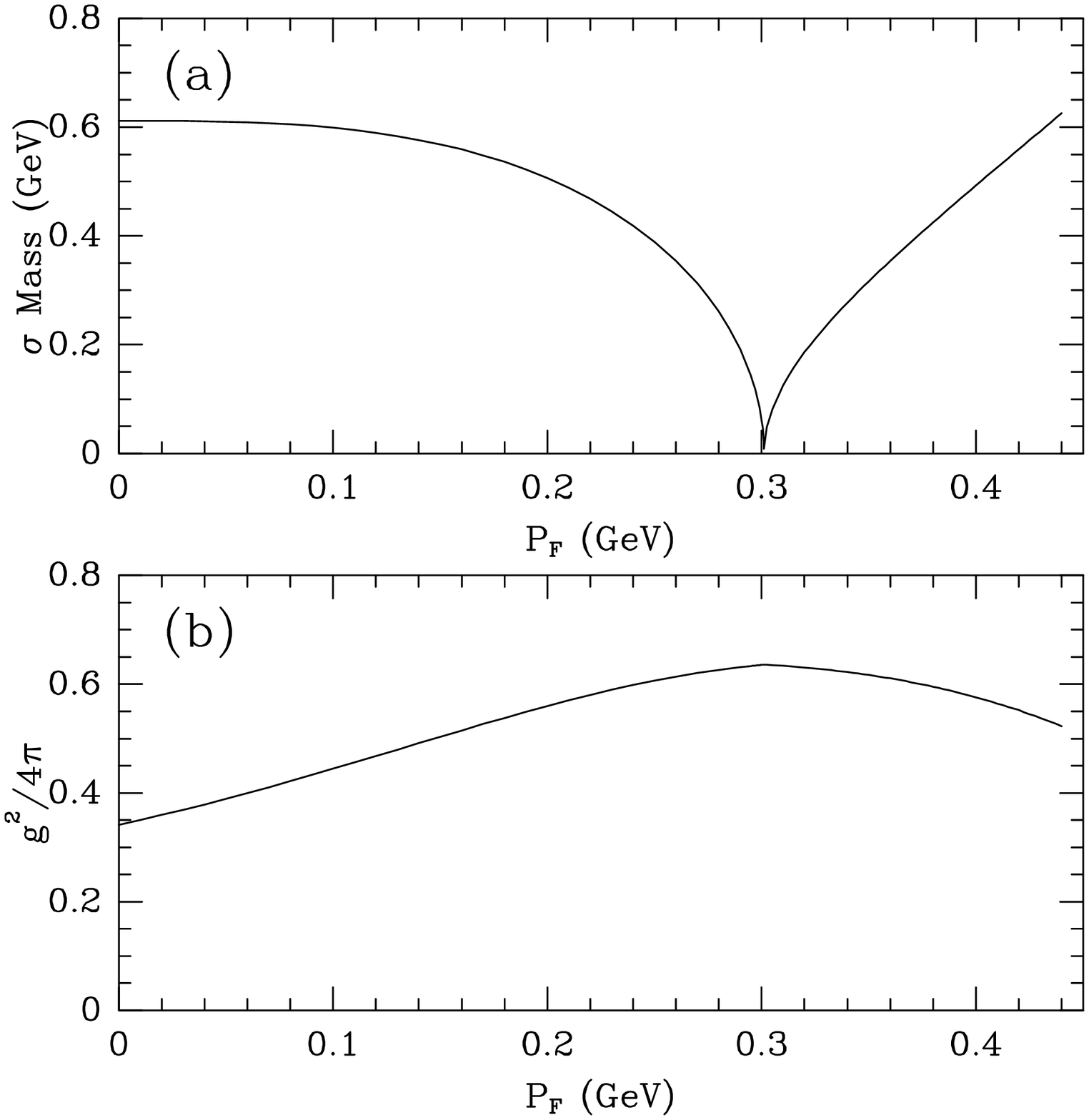,height=10cm,width=8cm}
\vfill\eject
\centerline{Figure 2}
\psfig{figure=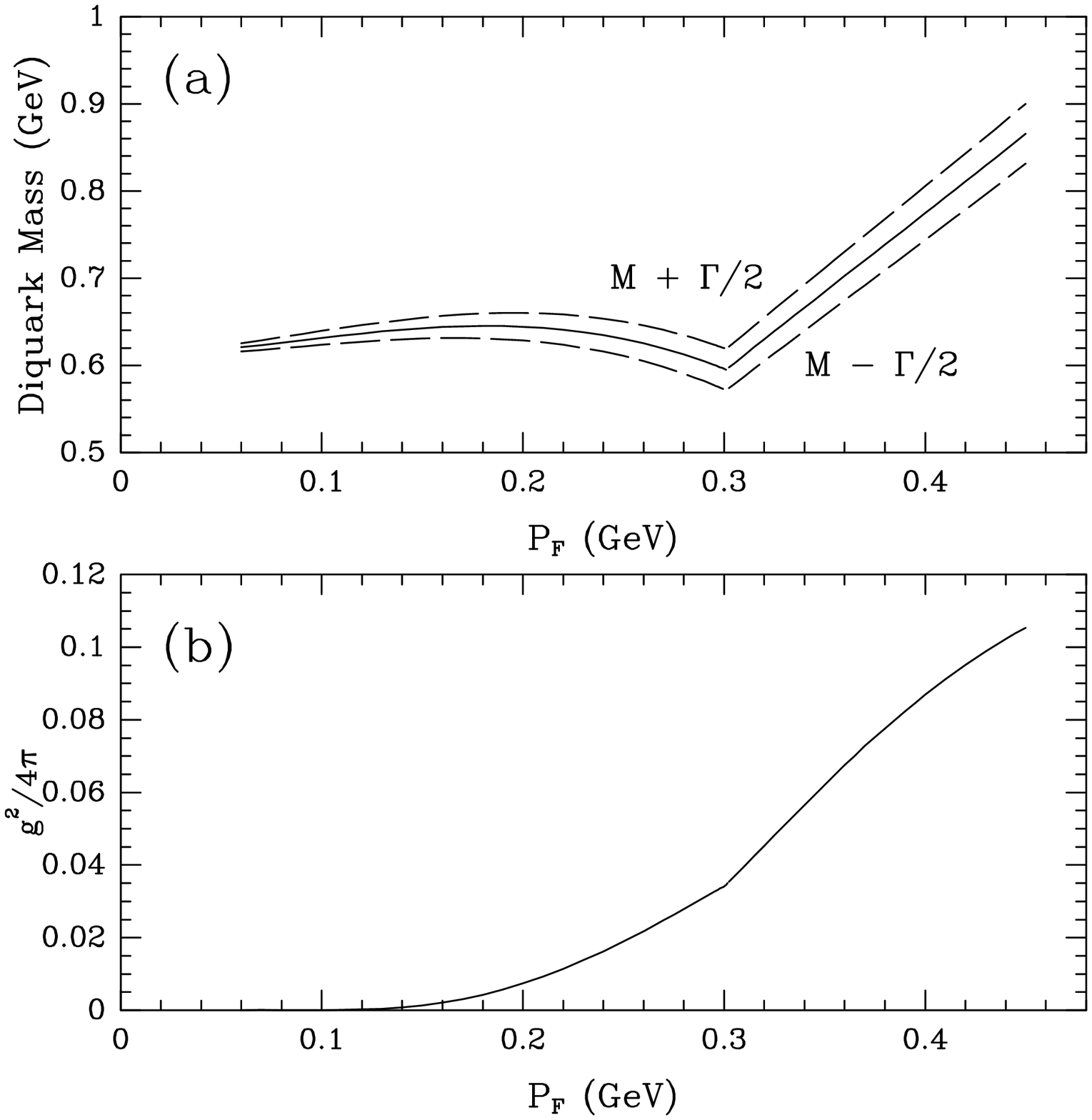,height=10cm,width=8cm}
\end{document}